\documentclass[sigconf]{acmart}

\usepackage{todonotes}
\usepackage{multirow}
\usepackage{colortbl}
\usepackage{xcolor}
\usepackage{listings} 
\usepackage{enumitem}
\usepackage{graphicx}

\AtBeginDocument{%
  \providecommand\BibTeX{{%
    \normalfont B\kern-0.5em{\scshape i\kern-0.25em b}\kern-0.8em\TeX}}}

\setcopyright{acmcopyright}
\copyrightyear{2024}
\acmYear{2024}
\acmDOI{XXXXXXX.XXXXXXX}

\acmConference[ACE 2024]{Australasian Computing Education Conference}{January,
  2024}{Sydney, Australia}
\acmPrice{15.00}
\acmISBN{978-1-4503-XXXX-X/18/06}

\begin{document}


\title{Decoding Logic Errors: A Comparative Study on Bug Detection by Students and Large Language Models}



\author{Stephen	MacNeil}
\affiliation{
  \institution{Temple University}
  \city{Philadelphia}
  \state{PA}
  \country{United States}}
\email{stephen.macneil@temple.edu}
\orcid{0000-0003-2781-6619}

\author{Paul Denny}
\affiliation{
  \institution{University of Auckland}
  \city{Auckland}
  \country{New Zealand}}
\email{paul@cs.auckland.ac.nz}
\orcid{0000-0002-5150-9806}

\author{Andrew Tran}
\affiliation{
  \institution{Temple University}
  \city{Philadelphia}
  \state{PA}
  \country{United States}}
\email{andrew.tran10@temple.edu}
\orcid{0000-0002-0094-1113}

\author{Juho Leinonen}
\affiliation{
  \institution{University of Auckland}
  \city{Auckland}
  \country{New Zealand}}
\email{juho.leinonen@auckland.ac.nz}
\orcid{0000-0001-6829-9449}

\author{Seth Bernstein}
\affiliation{
  \institution{Temple University}
  \city{Philadelphia}
  \state{PA}
  \country{United States}}
\email{seth.bernstein@temple.edu}
\orcid{0000-0001-5767-1057}

\author{Arto Hellas}
\affiliation{
  \institution{Aalto University}
  \city{Espoo}
  \country{Finland}}
\email{arto.hellas@aalto.fi}
\orcid{0000-0001-6502-209X}

\author{Sami Sarsa}
\affiliation{
  \institution{Aalto University}
  \city{Espoo}
  \country{Finland}}
\email{sami.sarsa@aalto.fi}
\orcid{0000-0002-7277-9282}

\author{Joanne Kim}
\affiliation{
  \institution{Temple University}
  \city{Philadelphia}
  \state{PA}
  \country{United States}}
\email{joanne.kim@temple.edu}
\orcid{0000-0001-7646-2373}

\renewcommand{\shortauthors}{Anon, et al.}

\begin{abstract}


Identifying and resolving logic errors can be one of the most frustrating challenges for novices programmers.  Unlike syntax errors, for which a compiler or interpreter can issue a message, logic errors can be subtle. In certain conditions, buggy code may even exhibit correct behavior -- in other cases, the issue might be about how a problem statement has been interpreted. Such errors can be hard to spot when reading the code, and they can also at times be missed by automated tests. 
There is great educational potential in automatically detecting logic errors, especially when paired with suitable feedback for novices.  Large language models (LLMs) have recently demonstrated surprising performance for a range of computing tasks, including generating and explaining code. 
These capabilities are closely linked to code syntax, which aligns with the next token prediction behavior of LLMs.  On the other hand, logic errors relate to the runtime performance of code and thus may not be as well suited to analysis by LLMs. To explore this, we investigate the performance of two popular LLMs, GPT-3 and GPT-4, for detecting and providing a novice-friendly explanation of logic errors. 
We compare LLM performance with a large cohort of introductory computing students $(n=964)$ solving the same error detection task.  Through a mixed-methods analysis of student and model responses, we observe significant improvement in logic error identification between the previous and current generation of LLMs, and find that both LLM generations significantly outperform students.
We outline how such models could be integrated into computing education tools, and discuss their potential for supporting students when learning programming.

\end{abstract}





\maketitle

\section{Introduction}

Learning to program involves navigating a landscape where mistakes are an inherent part of the journey. Novice programmers are bound to encounter numerous errors when writing code, ranging from logic flaws and syntactical inaccuracies to runtime glitches. These mistakes pose substantial hurdles to students as they strive to develop their programming skills. Despite extensive efforts by computing education researchers and practitioners to establish taxonomies and recognize patterns of common programming errors~\cite{brown2014investigating, mccall2019new, alqadi2017empirical, winikoff2014novice, malysheva2020bugs}
, the process of effectively detecting and resolving bugs remains a persistent challenge.

Simultaneously, the emergence of large language models (LLMs) has demonstrated remarkable capabilities in understanding and generating text that is highly similar to the text generated by people. These models, trained on vast amounts of textual data, have been used in a variety of computing education contexts including helping students to understand code~\cite{macneil2023experiences, leinonen2023comparing} and programming error messages~\cite{leinonen2023using}. These use cases demonstrate the ability of LLMs to understand the syntax and structure of code. Still, it is unclear whether models can reason about runtime performance without explicitly running the code. Therefore, detecting runtime errors may present a challenge for LLMs, limiting their potential to help learners. 

In this paper, we conduct a large-scale comparative study that investigates the abilities of two LLMs and students to detect bugs in faulty code. We recruited 964 students in a large introductory C programming class to identify bugs in three code examples. The selected code examples contained three types of bugs including an out-of-bounds error, an expression error, and an operator error. Students were selected because they are increasingly relying on LLMs as a legitimate help-seeking resource~\cite{zastudil2023generative, kabir2023answers, hellas2023exploring}. Our results suggest that LLMs outperform students in bug detection performance, especially for faulty code. However, in addition to detecting the pre-inserted bugs, the LLMs had a tendency to be overly proactive, also commenting on extremely minor `bugs' such as naming conventions, and other considerations that might be overwhelming if used for learning purposes. GPT-4 was nearly perfect at identifying bugs in faulty code, but was much more likely than GPT-3 to identify these minor `bugs' in the correct programs and therefore performed `poorly' on correct code. Studying correct code was important because students may use these tools when their code is mostly correct, and a list of minor errors may be demoralizing or may lead them off-track. Based on our findings, we conclude that LLMs appear to be capable of identifying logic errors, outperforming students at this task. However, additional work is needed to extend this work toward more complex code examples and with more advanced computing students. Given that experts are more likely to `chunk' code and see emergent structure, it is unclear whether they would be more or less able to identify bugs in the code without writing test cases. 



\section{Related Work}

\subsection{Students and Bugs in Code}

Bugs and errors are a common feature in student code and understanding the encountered problems and errors has been a long-standing endeavour within Computing Education Research. Early research in this area centered often on specific problems such as the looping problem or the rainfall problem~\cite{soloway1982what,soloway1983cognitive, johnson1983bug, seppala2015we}, leading also toward investigations into the design and features of programming languages (e.g.~\cite{soloway1983cognitive}). In general there are differences in frequency of programming errors~\cite{spohrer1986novice} and the time that it takes to fix those errors~\cite{denny2012all, brown2017novice, smith2019error, mccall2019new}. The types of errors that students encounter also gradually change~\cite{altadmri201537}, and they can stem from multiple sources~\cite{altadmri201537,ettles2018common}. These sources include misinterpreting the programming problem and having flaws in programming knowledge~\cite{ettles2018common}, not to mention the role of the used programming language~\cite{kohn2019error}. 

When students encounter a problem, they need to resolve it. Resolving programming problems -- or debugging -- can be done using multiple approaches, including tracing code, commenting out code, and adding print statements~\cite{murphy2008debugging,fitzgerald2008debugging,whalley2021novice}. Simply looking at the code and trying to find places that do not look right -- i.e. pattern matching -- can also be a viable strategy in some cases~\cite{fitzgerald2008debugging}. Like programming, finding problems in code by tracing the code is a skill, and both of them have been highlighted as something that students can struggle with. As an example, an ITiCSE working group from 2001 highlighted a lack of programming skills at the end of introductory programming course~\cite{mccracken2001multi}, and a subsequent ITiCSE working group from 2004 focused on the results by looking into students' ability to read and trace code~\cite{lister2004multi}, also highlighting problems. These issues have in part led to national and international efforts in understanding the struggles that students face, such as the BRACElet project that started in 2004~\cite{whalley2007many}. 

These studies tend to highlight that students have difficulties with tracing code~\cite{lister2004multi, vainio2007factors}, which might in part be explainable by lack of expertise. A student might, when solving a tracing problem, even just guess a solution if they do not have a higher-level reasoning strategy~\cite{lister2004multi}, or might simply have misconceptions about how a program executes, which in turn leads to faulty conclusions~\cite{vainio2007factors}. This possibility of guessing code tracing outcomes has also in part led to the emergence of ``explain in plain English'' problems. For these problems, students are expected to provide a high-level overview of the program's functionality and purpose rather than simply outlining what the program does~\cite{whalley2006australasian,lister2009further}. These problems can also be challenging, and any tools that would help students learn to understand and explain code would be of benefit.

\subsection{Generative AI and Computing Education} 


Recently computing education researchers are expressing concern and excitement about the ways that generative models may affect the computing education landscape~\cite{macneil2023implications, macneil2022automatically, prather2023transformed, prather2023robots, lau2023from, zastudil2023generative}. While a strong consensus about how we should adapt our pedagogical practice has yet to emerge, each of these discussions acknowledge that generative models are not likely a passing fad. 

Numerous examples of the capabilities of generative models are emerging such as their ability to both solve and create programming assignments~\cite{sarsa2022automatic, Finnie-Ansley2022Robots}, explain code~\cite{macneil2023experiences, leinonen2023comparing}, identify programming concepts~\cite{tran2023using}, answer multiple choice questions~\cite{savelka2023can, savelka2023large}, write code~\cite{puryear2022github, wermelinger2023using}, solve visual problems~\cite{hou2023more}, and enhance programming error messages~\cite{leinonen2023using}. These use cases are critical because without understanding the capabilities of generative models, it is extremely challenging to adapt to this rapidly changing landscape.

However, limited work has investigated the capabilities of generative models to identify bugs within code. Given that novice programmers often encounter bugs and may lack the ability to identify and fix these bugs, it is important to explore the capabilities of generative models to accomplish this task. Very recent papers focus on enhancing programming error messages~\cite{leinonen2023using} and automatically repairing bugs in code~\cite{koutcheme2023automated, fan2023automated, jiang2023impact}. In this paper, we add to the growing set of use cases by exploring the potential for generative models to identify potential bugs and errors.

\section{Method}

\begin{figure*}
    \centering
    \includegraphics[width=\linewidth]{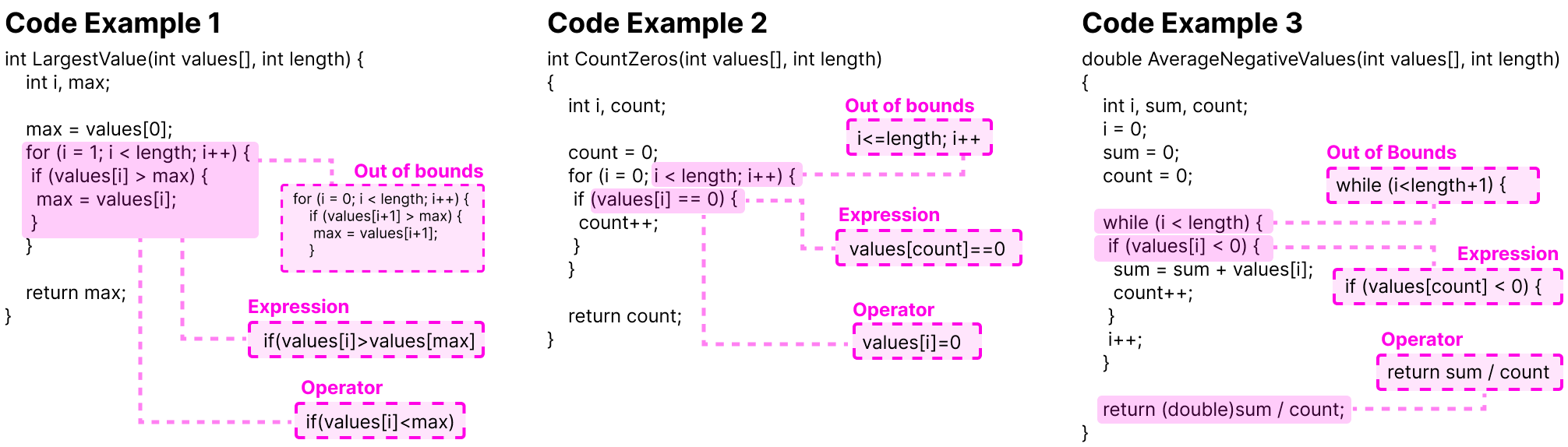}
    \caption{The three examples of code with the correct variant, and the three incorrect variants annotated. The incorrect variants were 1) operator error, 2) expression error, and 3) out-of-bounds error. }
    \label{fig:three-code-snippets}
\end{figure*}

\subsection{Research Questions}

Previous research has demonstrated many impressive capabilities of large language models. However, many of these examples, such as generating explanations and identifying programming concepts, are closely linked to code syntax, which aligns with the next token prediction behavior of LLMs. To better explore the potential limits of LLMs, this study focuses on identifying logic errors in code, which relate to the runtime performance of code, and thus may not be as well suited to analysis by LLMs as they are unable to execute code. If large language models perform well in this task, there is an exciting opportunity to use these models to help students to debug their code. Based on these goals, we investigated the following research questions: 

\begin{enumerate}
    \item [\textbf{RQ 1:}] How do students and large language models compare in their ability to correctly identify logic errors in faulty code?
    \item [\textbf{RQ 2:}] Which types of logic errors are easiest for students and large language models to correctly identify? 
    \item [\textbf{RQ 3:}] How many bugs or issues do students and large language models identify when reviewing faulty and correct code?
\end{enumerate}

\subsection{Study Design}

In this study, we seek to investigate the performance of large language models in detecting bugs in faulty code. We conducted a study that compared the performance of students with the two large language models GPT-3 and GPT-4. Performance was measured across three code examples with four variants. These variants included the correct code and three variants with bugs introduced: 1) an operator error, 2) an out-of-bounds error, and 3) an expression error. The study was designed with two between-subjects components which include the source of the detection method, i.e., whether it was performed by the students, GPT-3, or GPT-4, and the bug variant. The study also included a within-subjects component which was the three code examples. By showing students multiple examples, we could partially control for participant error. 

\subsubsection{Participants, Data Collection, and Ethics} 

The data used in this study were collected from a first-year C programming course at The University of Anonymous. The data were collected during a single lab session that ran over a one-week period. Leading up to this lab, the course covered the concepts of arithmetic, types, functions, loops, and arrays. We collected 964 total complete responses from students. The data collection followed the ethical guidelines of the university and was approved by the ethics review board\footnote{IRB approval number anonymized for review.}.

\subsubsection{Study Tasks}

As part of the lab, students were shown three code examples. Figure~\ref{fig:three-code-snippets} shows the three examples that were shown to students during the lab. Each example contains a function with a single loop that processes elements of an array.  The task for the students was to identify any bugs that might exist within the code. The instructions said \textit{``Consider the following definition of a function called <Function Name>:''} which was followed by the code without comments. They were then asked to come up with a short description of what they believe the intended purpose of the function to be. This was followed by having them \textit{``List all errors, if any, found in this code based on your explanation of the purpose of the function. It is possible that the code contains one or more small errors (however, this is not necessarily true and the code may be correct). If you can identify any errors in the implementation of the code, you should describe these errors.''} 

\subsubsection{Measures}

The data collection resulted in 2980 total responses from students. In addition, 30 LLM responses were generated for each code example and version pair by varying the temperature and prompt to account for variations that might affect performance. This resulted in 720 total additional responses from the two models. 

A team of four researchers manually coded each student and model response. The coders evaluated the \textbf{correctness} of the identified bug as a dichotomous variable (e.g.: correct or incorrect). The coders also evaluated the \textbf{number of bugs} that the response contained. The coding was mutually exclusive: a response correctly identifying a bug but also noting other incorrect bugs was coded as correct. When coding the example that did not contain bugs, we coded a blank response or an explicit statement that no bugs were contained as a correct response and other responses were considered incorrect. This coding scheme did not allow for explicitly tracking false positives and false negatives, but it was necessary to obtain substantial inter-rater reliability ($\kappa=0.873$, 30 ratings). Students often did not explicitly state the bug so we coded their response as `correct' even if they only provided a solution that would fix the expected bug. 

\subsubsection{Analysis for Conditional Differences} 

We analyzed the dependent measures (e.g.: number of bugs) using a linear mixed-effects model. The main fixed factors of interest were the ``Source'' (representing GPT-3, GPT-4, or Students) and the ``Version'' of the code example (representing different versions of the example). Additionally, an interaction term between ``Source'' and ``Version'' was included to examine potential differences in bug identification across sources and versions. To account for potential dependencies among observations from the same example, a random intercept term was included in the model specification. This random effect was nested within the ``Code Example'' factor, capturing the variability associated with different examples. Pairwise comparisons were made using the Tukey method with Holm's correction for multiple comparisons.

\subsection{Models}

\subsubsection{Model Specification}

To automatically identify the bugs in the study, we used two large language models~\cite{brown2020language} developed by OpenAI. The first model, \textit{text-davinci-003}, has been widely used up until the time of running the study. Later, when GPT-4 was released, we included results using the \textit{gpt-4-0314} model to understand how the state-of-the-art models perform at the same task. 

\subsubsection{Prompt Engineering}

Prompt engineering is a process of developing instructions to guide the responses of an LLM. The specificity and phrasing of these prompts have the potential to strongly influence the content and quality of the responses~\cite{Arora2022AskMA, Zhao2021CalibrateBU, si2023prompting}. Understanding the potential effects that prompts can have on performance, we used multiple prompting strategies to account for this aspect. In addition, the hyperparameters of an LLM, such as the temperature, can also affect the output. Lower temperatures tend to result in more deterministic responses while higher temperatures tend to provide more `creative' responses. We chose to use the default temperature of $0.7$ and a lower temperature of $0.3$. The three prompts used for this study are listed below. 

\begin{itemize}  
    \item \# List all errors and bugs, if any, found in the following C code: <code>
    \item \# List any issues, including bugs, errors, or potential problems that exist in the following C code: <code>
    \item \# Assume the role of a highly intelligent computer scientist who is capable of easily finding bugs and errors by reading source code. List all errors and bugs, if any, found in the following C code: <code> 
\end{itemize}

Between the variations in prompt and temperature, there were 6 possible permutation. For each permutation, we issued 5 requests to the OpenAI API. The reason for issuing 5 requests was to account for the non-deterministic nature of LLM prompts. This resulted in 30 responses for each combination of code example and bug type and 360 total requests to OpenAI.




\section{Results}


\begin{table*}
\centering
\caption{A summary of student and model performance in correctly identifying bugs. In instances where coders were unsure, they coded it as `uncertain.' These are excluded from the table. For instance, if we consider the number of correct and incorrect responses for GPT-3 in Code Example 1 with Bug 2, their total does not sum to 30.}
\begin{tabular}{|cr|c|c|c|c|c|c|c|c|c|c|c|c|}
\hline
\multicolumn{2}{|c|}{} & \multicolumn{4}{c|}{Code Example 1} & \multicolumn{4}{c|}{Code Example 2} & \multicolumn{4}{c|}{Code Example 3} \\
\cline{1-14}
 Source &  & Bug 1 & Bug 2 & Bug 3 & Correct & Bug 1 & Bug 2 & Bug 3 & Correct & Bug 1 & Bug 2 & Bug 3 & Correct \\
\hline
\multirow{3}{*}{Student} & \cellcolor{blue!10}correct & \cellcolor{blue!10}56 & \cellcolor{blue!10}91 & \cellcolor{blue!10}108 & \cellcolor{blue!10}207 & \cellcolor{blue!10}71 & \cellcolor{blue!10}90 & \cellcolor{blue!10}82 & \cellcolor{blue!10}222 & \cellcolor{blue!10}84 & \cellcolor{blue!10}66 & \cellcolor{blue!10}69 & \cellcolor{blue!10}220 \\
& \cellcolor{red!10}incorrect & \cellcolor{red!10}147 & \cellcolor{red!10}146 & \cellcolor{red!10}125 & \cellcolor{red!10}19 & \cellcolor{red!10}184 & \cellcolor{red!10}165 & \cellcolor{red!10}130 & \cellcolor{red!10}6 & \cellcolor{red!10}155 & \cellcolor{red!10}143 & \cellcolor{red!10}162 & \cellcolor{red!10}26 \\
& rate & 0.276 & 0.384 & 0.464 & 0.916 & 0.278 & 0.353 & 0.387 & 0.974 & 0.351 & 0.316 & 0.299 & 0.894 \\
\hline
\multirow{3}{*}{GPT-3} & \cellcolor{blue!10}correct & \cellcolor{blue!10}30 & \cellcolor{blue!10}23 & \cellcolor{blue!10}8 & \cellcolor{blue!10}30 & \cellcolor{blue!10}29 & \cellcolor{blue!10}27 & \cellcolor{blue!10}24 & \cellcolor{blue!10}29 & \cellcolor{blue!10}29 & \cellcolor{blue!10}30 & \cellcolor{blue!10}24 & \cellcolor{blue!10}12 \\
& \cellcolor{red!10}incorrect & \cellcolor{red!10}0 & \cellcolor{red!10}4 & \cellcolor{red!10}22 & \cellcolor{red!10}0 & \cellcolor{red!10}0 & \cellcolor{red!10}0 & \cellcolor{red!10}1 & \cellcolor{red!10}1 & \cellcolor{red!10}0 & \cellcolor{red!10}0 & \cellcolor{red!10}6 & \cellcolor{red!10}17 \\
& rate & 1 & 0.852 & 0.267 & 1 & 1 & 1 & 0.96 & 0.967 & 1 & 1 & 0.800 & 0.414 \\
\hline
\multirow{3}{*}{GPT-4} & \cellcolor{blue!10}correct & \cellcolor{blue!10}29 & \cellcolor{blue!10}30 & \cellcolor{blue!10}28 & \cellcolor{blue!10}19 & \cellcolor{blue!10}30 & \cellcolor{blue!10}30 & \cellcolor{blue!10}28 & \cellcolor{blue!10}0 & \cellcolor{blue!10}30 & \cellcolor{blue!10}30 & \cellcolor{blue!10}28 & \cellcolor{blue!10}0 \\
& \cellcolor{red!10}incorrect & \cellcolor{red!10}0 & \cellcolor{red!10}0 & \cellcolor{red!10}1 & \cellcolor{red!10}11 & \cellcolor{red!10}0 & \cellcolor{red!10}0 & \cellcolor{red!10}1 & \cellcolor{red!10}30 & \cellcolor{red!10}0 & \cellcolor{red!10}0 & \cellcolor{red!10}1 & \cellcolor{red!10}30 \\
& rate & 1 & 1 & 0.966 & 0.633 & 1 & 1 & 0.966 & 0 & 1 & 1 & 0.966 & 0 \\
\hline
\end{tabular}
\label{tab:performance}
\end{table*}

\subsection{Bug Detection Performance}

Performance in bug detection rates varied between the students and the models, as shown in Table~\ref{tab:performance}. GPT-3 exhibited an overall correctness rate of 85.3\%, while GPT-4 closely followed with a correctness rate of 85.0\%. Notably, students had a much lower bug detection rate at 49.1\%. While both models detected bugs at nearly twice the rate of students, performance was even higher when only considering model performance on faulty code. 

\subsubsection{For faulty code, LLMs outperform students}

When presented with incorrect code, GPT-3 exhibited a bug detection rate of 87.3\%, demonstrating a substantial ability to identify coding errors. GPT-4 surpassed this performance with an impressive bug detection rate of 99.2\%, indicating a higher sensitivity to identifying bugs within faulty code. On the other hand, students detected bugs at rate of 34.5\%, showcasing a limited proficiency in detecting coding errors. 


\subsubsection{LLMs tended to identify bugs in correct code}

In the case of identifying correctly functioning code, GPT-3 achieved a bug detection rate of 79.4\% (i.e., classified the code as bug-free). GPT-4, however, displayed a comparatively lower rate of 42.2\% in correctly identifying bug-free code. In contrast, students demonstrated a notably high proficiency in identifying correct code, with a bug detection rate of 92.8\%. 

\subsection{Number of Bugs Detected}

We observed statistically significant differences in the number of bugs identified by GPT-3, GPT-4, and students. The results of the linear mixed-effects model, which are summarized in Table~\ref{tab:bug-count}, show that GPT-4 identified significantly more bugs than GPT-3 ($\beta = 0.76$, $SE = 0.13$, $z = 5.77$, $p < 0.001$) and students ($\beta = 1.70$, $SE = 0.11$, $z = -15.13$, $p < 0.001$). The model estimated that GPT-4 identified 0.761 more bugs than GPT-3 and 1.701 more bugs than students when other variables were held constant. GPT-3 also identified statistically significantly more bugs than students ($\beta = 0.94$, $SE = 0.11$, $z = -8.30$, $p < 0.001$). 

\begin{table}
\centering
\caption{Number of bugs detected by condition}
\begin{tabular}{lccccc}
\hline
Contrasts & Estimate & Std. Error & z value & \text{Pr(>|z|)} \\
\hline
GPT4 - GPT3 & 0.761 & 0.132 & 5.766 & $< 0.001$ *** \\
GPT3 - Student & 0.940 & 0.113 & 8.303 & $< 0.001$ *** \\
GPT4 - Student  & 1.701 & 0.112 & 15.133 & $< 0.001$ *** \\
Bug2 - Bug1  & 0.390 & 0.134 & 2.910 & 0.02724 * \\
Correct - Bug2 & 0.652 & 0.186 & 3.495 & 0.00389 ** \\
Bug3 - Bug1  & 0.644 & 0.139 & 4.626 & $< 0.001$ *** \\
Correct - Bug2  & 0.262 & 0.187 & 1.399 & 0.63514 \\
Bug3 - Bug2 & 0.254 & 0.140 & 1.812 & 0.36131 \\
Bug3 - Correct & -0.008 & 0.190 & -0.040 & 1.00000 \\
\hline
\end{tabular}
\label{tab:bug-count}
\footnotesize{\\ \textit{Signif. codes:} 0 ‘***’ 0.001 ‘**’ 0.01 ‘*’ 0.05 ‘.’ 0.1 ‘ ’ 1 \\ (Adjusted p values reported -- Holm Correction)}
\end{table}

\subsection{Analyzing the Bug Reports}

\subsubsection{GPT-4 was more verbose, even when normalized by the number of bugs detected} 

We computed the average word count for responses made by students and each model. GPT-4 responses had on average 129.0 ($\sigma = 44.7$) words followed by GPT-3 and students with 54.2 ($\sigma = 19.5$) and 38.9 ($\sigma = 27.0$) words respectively. This constitutes a 3.31-fold increase in the number of words GPT-4 produced compared to students. Given the differences in number of bugs identified by source, we normalized word count by the number of bugs reported. This resulted in 52.7 ($\sigma = 25.7$) words for GPT-4 and 23.5 ($\sigma = 9.11$) and 35.6 ($\sigma = 24.5$) words for GPT-3 and students. These results should be contextualized by the observation that GPT-4 had a tendency to provide partial and in a few cases complete solutions for the bugs that it identified. Moreover, the models exhibited a more concentrated distribution around their means, while student responses exhibited notably higher variability. 


\subsubsection{Qualitative analysis of responses}

To better understand the capabilities of large language models in detecting bugs in code and to draw distinctions between students' responses and these models, we did a qualitative exploration of the error messages. By analyzing selected bug reports generated by the models in our study, we could shed light on a variety of shortcomings of the models' approaches. As suggested by our analysis of word count, there appear to be differences in the ways that LLMs and students identify and describe bugs. Students tended to describe the bug, offer a solution, or both. We did not observe an instance where a model offered a solution without also describing the underlying bug. Models often combined a description of the bug along with a partial solution, in some cases they provided a complete solution. GPT-4 appeared to be most likely to offer a complete solution. As noted in the analysis of bug detection performance, LLMs often identified bugs even for `correct' code. Below is an example where GPT-4 identified 6 bugs in otherwise correct code for Example 2: 

\begin{enumerate}
    \item No function prototype: There is no function prototype provided for the \texttt{CountZeros} function. It is a good practice to provide a function prototype before the function definition to avoid any potential issues with implicit declarations.
    \item No include statements: The code does not include any necessary header files, such as \texttt{stdio.h} or \texttt{stdlib.h}, which might be required for certain operations or functions used in the program.
    \item Inconsistent naming convention: The function name \\\texttt{COUNTZEROS} is mentioned in the question, but the actual function name in the code is \texttt{CountZeros}. It is important to maintain a consistent naming convention throughout the code.
    \item No error handling: The function does not perform any error handling, such as checking for NULL pointers or negative array length. This could lead to undefined behavior if the function is called with invalid arguments.
    \item No comments: The code lacks comments explaining the purpose of the function and its parameters. Adding comments can make the code more readable and easier to understand for others.
    \item Potential integer overflow: If the \texttt{length} parameter is very large
\end{enumerate}

It should be noted that most of these identified `bugs' could be interpreted as areas for improvement. For example, the model suggests improvements to naming conventions or including comments. However, both the fourth and sixth bugs could actually result in runtime issues. The model identifies a lack of input sanitization checks for two potential corner cases. The qualitative investigation underscores the intricacies of LLMs' code interpretations. They exhibit detailed understanding but can occasionally miss the mark on context or offer feedback that, while technically correct, might feel misaligned with the coder's intent. Balancing LLM insights with human discernment could yield the most effective outcomes. 



Many student responses just contained the proposed change without any explanation or reasoning. They often did not explicitly define a bug but instead only described the solution. Some students also indicated errors that either did not fix the issue, introduced new problems, or focused too heavily on syntactical correctness without addressing the core problem. In the example below, a student highlighted changes that should be made to the code which do not fix the bug:

\begin{itemize}
    \item \texttt{i = 0;} should change to \texttt{i = 1;} to avoid using the 0th value.
    \item Instead of \texttt{count++}, use \texttt{count = count + 1;}.
    \item There should be no space between \texttt{for} and the opening parenthesis \texttt{(}.
    \item Similarly, there should be no space between \texttt{if} and the opening parenthesis \texttt{(}.
\end{itemize}



\section{Discussion}

Our results suggest that large language models are more capable than students at identifying bugs in code. There are multiple possible explanations for this. First, more expert programmers often do not necessarily need to read the code character by character or word by word when forming an understanding of the code, rather, they study features of the code that are relevant to the task at hand~\cite{heinonen2023synthesizing}. Consequently, a student may miss syntax errors or minor bugs, if they are not in focus. This can also be explained by the happy path mentality where because most of the code is correct, students may become complacent and fail to detect bugs; some bugs also take more time to identify and fix than others~\cite{denny2012all}. Participants were explicitly prompted to find errors, which puts them into an explicit debugging mindset. In practice, they might not critically examine their code with the same scrutiny, so the bug detection rate for students may actually be even lower in practice.


Both LLMs performed extremely well, with GPT-4 performing near perfect when presented with buggy code. However, both models performed poorly in our analysis of correct code as they identified very minor bugs and stylistic aspects such as naming conventions contrary to our expectation that they would classify the code as bug-free. While the suggestions were largely correct, it might not be helpful to point out minor bugs and code conventions in otherwise correct code, especially considering students' preferences for concise bug reports~\cite{denny2021designing}.  

One noticeable difference between GPT-3 and GPT-4 was that GPT-4 would point out these minor bugs more than GPT-3. One possible explanation for this is that the newer model has possibly had more instruction fine-tuning, where the model is trained to follow instructions from the user. This might cause the model to try please the user by going above and beyond the ask, e.g. in our case not only pointing out the obvious bug, but also commenting on more minor issues. We also found that GPT-4 was more verbose, even when controlling for the number of bugs in the code. This aligns with prior findings where newer models often add superfluous textual content to responses~\cite{denny2023promptly} and may come up with non-existing bugs to fix when asked to help with buggy code~\cite{hellas2023exploring}. 

The ability of LLMs to correctly identify bugs at a much higher rate than students has exciting implications for computing education. LLMs could be used to help novices (and more experienced programmers too) in detecting bugs in code, for example, by having LLMs integrated directly into the IDE that students use to work on their course exercises. Models could make suggestions for improvement as they did in cases with correct code or identify subtle logic errors in the code, potentially building on prior research on improving programming error messages, which has the promise of improving learning~\cite{becker2019compiler, denny2021designing}. Despite the allure of the technological possibilities, there likely should be a mechanism that would control how often the suggestions would be shown, as not all errors require help~\cite{hellas2023exploring}. Similarly, it is important to carefully curate educational content, especially with growing concerns about over-reliance on LLMs~\cite{lau2023from, denny2023conversing, zastudil2023generative, macneil2023implications, wermelinger2023using}. To mitigate potential issues, it is likely preferable to avoid directly presenting errors and solutions to students. Instead, pedagogical systems could detect when students are spinning their wheels trying to debug their code~\cite{beck2013wheel} and then use the LLM to scaffold students toward identifying the error themselves. Thus providing learning opportunities that also mitigate stress associated with debugging. 

Similarly, as LLMs are adept at detecting bugs and writing suggestions on how to fix them, they could be further integrated into teacher tools. As an example, tools such as OverCode~\cite{glassman2015overcode} and CodeClusters~\cite{koivisto2022evaluating} that are designed to provide feedback to masses of students could be integrated with LLMs so that LLMs would create draft feedback, which instructors then could -- when needed -- adjust and send out. The ability of LLMs to identify rare corner cases also has interesting implications for teaching testing, as feedback from LLMs could help with writing more comprehensive test suites. The good performance of the models could also lead to new, innovative exercise types. For example, we envision that an LLM could create buggy code where students would need to find and fix the bug -- similarly, one activity could be trying to create bugs that LLMs fail to identify. Such activities could also provide additional data on learning, which then could be used to fine-tune LLMs.

As the educational landscape continues to adapt to LLMs~\cite{prather2023robots, prather2023transformed, zastudil2023generative, lau2023from}, the new bug capabilities of LLMs identified in this paper may further inform how students seek help in classroom settings~\cite{hou2023effects}.

\subsection{Limitations}

To make the task more ecologically valid, we provided students with an open-response question rather than a multiple-choice question. This had the advantage that students could not guess the right answer and was more similar to how students would encounter code in the wild; however, it became difficult to differentiate between a response that explicitly stated `no bugs' and a blank response. To address this limitation, we evaluated the rates of default responses by variant and observed no statistically significant difference in the number of default responses across all four variants.

Participants were asked to identify any bugs that were present within the code, so in this case, a lack of an explicit response was treated as a default response (e.g.: `no bugs'). To assess the impact on our results, we recalculated percentages by excluding blanks. The revised student correctness rates are as follows: 79.6\% (133 blanks removed), 89.8\% (169 blanks removed), and 60.0\% (181 blanks removed). These results represent a conservative estimate, considering only explicitly stated correct answers. The resulting rates remained higher than GPT-4, but closer to GPT-3 correctness rates. 

Participants were also explicitly instructed to identify bugs as part of the lab activity. While prior research has demonstrated that debugging others' code can be challenging~\cite{whalley2021novice}, it is possible that if students were studying their own code, it might have been easier for them. Relatedly, the code did not have comments that would explain what each line of code does. This may align with code students often encounter naturally, but could have affected the students' performance or required the model to infer too much from the code structure and function name. 

The code examples used in this study only contained a single intended error. It is possible that the presence of multiple bugs in code might affect the performance of LLMs (and students) in detecting bugs. The goal for this paper was an initial tightly scoped investigation of identifying a bug within code. Future work will investigate cases where multiple bugs are included. 

In our study, we employed a robust approach by utilizing three distinct prompts, leveraging multiple models, including both GPT-3 and GPT-4, and exploring various temperatures (i.e., 0.4 and 0.7). Additionally, each prompt was issued multiple times to accommodate the inherent probabilistic nature of generative AI. While we acknowledge the potential impact of further prompt optimization on mitigating false positives in the correct code condition, it's essential to note the dynamic nature of these models, characterized by continuous changes in verbosity and performance~\cite{savelka2023large, savelka2023thrilled, teebagy2023improved}. Rather than providing a definitive characterization of performance, our primary objective was to delve into a novel capability of LLMs.


\section{Conclusion}

In this work, we report on the results of a study that compares the ability of students and large language models to identify bugs in faulty and correct code. Our results suggest that students struggled to find bugs in faulty code, but that they performed relatively well at identifying whether the code was correct. The models performed in the opposite way: both models (GPT-3 and GPT-4) strongly outperformed students in identifying bugs in faulty code, but tended to identify many minor `bugs' which were more akin to suggestions for improvement when the code was correct. This suggests that models are overly sensitive toward discovering bugs in code. While some of the minor bugs detected by the models could be considered `bugs', such over-sensitivity could be seen as a negative for integrating LLMs into teaching. If students receive superflous feedback on minor stylistic aspects, for example, they might start disregarding any useful feedback from the models too.



\balance
\bibliographystyle{ACM-Reference-Format}
\bibliography{references,ref-dump-1,ref-dump-2}


\begin{thebibliography}{62}


\ifx \showCODEN    \undefined \def \showCODEN     #1{\unskip}     \fi
\ifx \showDOI      \undefined \def \showDOI       #1{#1}\fi
\ifx \showISBNx    \undefined \def \showISBNx     #1{\unskip}     \fi
\ifx \showISBNxiii \undefined \def \showISBNxiii  #1{\unskip}     \fi
\ifx \showISSN     \undefined \def \showISSN      #1{\unskip}     \fi
\ifx \showLCCN     \undefined \def \showLCCN      #1{\unskip}     \fi
\ifx \shownote     \undefined \def \shownote      #1{#1}          \fi
\ifx \showarticletitle \undefined \def \showarticletitle #1{#1}   \fi
\ifx \showURL      \undefined \def \showURL       {\relax}        \fi
\providecommand\bibfield[2]{#2}
\providecommand\bibinfo[2]{#2}
\providecommand\natexlab[1]{#1}
\providecommand\showeprint[2][]{arXiv:#2}

\bibitem[Alqadi and Maletic(2017)]%
        {alqadi2017empirical}
\bibfield{author}{\bibinfo{person}{Basma~S Alqadi} {and}
  \bibinfo{person}{Jonathan~I Maletic}.} \bibinfo{year}{2017}\natexlab{}.
\newblock \showarticletitle{An empirical study of debugging patterns among
  novices programmers}. In \bibinfo{booktitle}{\emph{Proc. of the 2017 ACM
  SIGCSE technical Symp. on computer science education}}.
  \bibinfo{pages}{15--20}.
\newblock


\bibitem[Altadmri and Brown(2015)]%
        {altadmri201537}
\bibfield{author}{\bibinfo{person}{Amjad Altadmri} {and}
  \bibinfo{person}{Neil~CC Brown}.} \bibinfo{year}{2015}\natexlab{}.
\newblock \showarticletitle{37 million compilations: Investigating novice
  programming mistakes in large-scale student data}. In
  \bibinfo{booktitle}{\emph{Proc. of the 46th ACM Technical Symp. on Computer
  Science Education}}. \bibinfo{pages}{522--527}.
\newblock


\bibitem[Arora et~al\mbox{.}(2022)]%
        {Arora2022AskMA}
\bibfield{author}{\bibinfo{person}{Simran Arora}, \bibinfo{person}{Avanika
  Narayan}, \bibinfo{person}{Mayee~F. Chen}, \bibinfo{person}{Laurel~J. Orr},
  \bibinfo{person}{Neel Guha}, \bibinfo{person}{Kush~S Bhatia},
  \bibinfo{person}{Ines Chami}, \bibinfo{person}{Frederic Sala}, {and}
  \bibinfo{person}{Christopher R'e}.} \bibinfo{year}{2022}\natexlab{}.
\newblock \showarticletitle{Ask Me Anything: A simple strategy for prompting
  language models}.
\newblock \bibinfo{journal}{\emph{ArXiv}}  \bibinfo{volume}{abs/2210.02441}
  (\bibinfo{year}{2022}).
\newblock


\bibitem[Beck and Gong(2013)]%
        {beck2013wheel}
\bibfield{author}{\bibinfo{person}{Joseph~E Beck} {and} \bibinfo{person}{Yue
  Gong}.} \bibinfo{year}{2013}\natexlab{}.
\newblock \showarticletitle{Wheel-spinning: Students who fail to master a
  skill}. In \bibinfo{booktitle}{\emph{Int. conf. on artificial intelligence in
  education}}. Springer, \bibinfo{pages}{431--440}.
\newblock


\bibitem[Becker et~al\mbox{.}(2019)]%
        {becker2019compiler}
\bibfield{author}{\bibinfo{person}{Brett~A Becker}, \bibinfo{person}{Paul
  Denny}, \bibinfo{person}{Raymond Pettit}, \bibinfo{person}{Durell Bouchard},
  \bibinfo{person}{Dennis~J Bouvier}, \bibinfo{person}{Brian Harrington},
  \bibinfo{person}{Amir Kamil}, \bibinfo{person}{Amey Karkare},
  \bibinfo{person}{Chris McDonald}, \bibinfo{person}{Peter-Michael Osera},
  {et~al\mbox{.}}} \bibinfo{year}{2019}\natexlab{}.
\newblock \showarticletitle{Compiler error messages considered unhelpful: The
  landscape of text-based programming error message research}.
\newblock \bibinfo{journal}{\emph{Proc. of the working group reports on
  innovation and technology in computer science education}}
  (\bibinfo{year}{2019}).
\newblock


\bibitem[Brown and Altadmri(2014)]%
        {brown2014investigating}
\bibfield{author}{\bibinfo{person}{Neil~CC Brown} {and} \bibinfo{person}{Amjad
  Altadmri}.} \bibinfo{year}{2014}\natexlab{}.
\newblock \showarticletitle{Investigating novice programming mistakes: Educator
  beliefs vs. student data}. In \bibinfo{booktitle}{\emph{Proc. of the tenth
  annual Conf. on Int. computing education research}}. \bibinfo{pages}{43--50}.
\newblock


\bibitem[Brown and Altadmri(2017)]%
        {brown2017novice}
\bibfield{author}{\bibinfo{person}{Neil~CC Brown} {and} \bibinfo{person}{Amjad
  Altadmri}.} \bibinfo{year}{2017}\natexlab{}.
\newblock \showarticletitle{Novice Java programming mistakes: Large-scale data
  vs. educator beliefs}.
\newblock \bibinfo{journal}{\emph{ACM Trans. on Computing Education (TOCE)}}
  \bibinfo{volume}{17}, \bibinfo{number}{2} (\bibinfo{year}{2017}),
  \bibinfo{pages}{1--21}.
\newblock


\bibitem[Brown et~al\mbox{.}(2020)]%
        {brown2020language}
\bibfield{author}{\bibinfo{person}{Tom Brown}, \bibinfo{person}{Benjamin Mann},
  \bibinfo{person}{Nick Ryder}, \bibinfo{person}{Melanie Subbiah},
  \bibinfo{person}{Jared~D Kaplan}, \bibinfo{person}{Prafulla Dhariwal},
  \bibinfo{person}{Arvind Neelakantan}, \bibinfo{person}{Pranav Shyam},
  \bibinfo{person}{Girish Sastry}, \bibinfo{person}{Amanda Askell},
  {et~al\mbox{.}}} \bibinfo{year}{2020}\natexlab{}.
\newblock \showarticletitle{Language models are few-shot learners}.
\newblock \bibinfo{journal}{\emph{Advances in neural information processing
  systems}}  \bibinfo{volume}{33} (\bibinfo{year}{2020}),
  \bibinfo{pages}{1877--1901}.
\newblock


\bibitem[Denny et~al\mbox{.}(2023a)]%
        {denny2023conversing}
\bibfield{author}{\bibinfo{person}{Paul Denny}, \bibinfo{person}{Viraj Kumar},
  {and} \bibinfo{person}{Nasser Giacaman}.} \bibinfo{year}{2023}\natexlab{a}.
\newblock \showarticletitle{Conversing with Copilot: Exploring Prompt
  Engineering for Solving CS1 Problems Using Natural Language}. In
  \bibinfo{booktitle}{\emph{Proc. of the 54th ACM Technical Symp. on Computer
  Science Education V. 1}} (Toronto ON, Canada) \emph{(\bibinfo{series}{SIGCSE
  2023})}. \bibinfo{publisher}{ACM}, \bibinfo{address}{New York, NY, USA},
  \bibinfo{pages}{1136–1142}.
\newblock
\showISBNx{9781450394314}
\urldef\tempurl%
\url{https://doi.org/10.1145/3545945.3569823}
\showDOI{\tempurl}


\bibitem[Denny et~al\mbox{.}(2023b)]%
        {denny2023promptly}
\bibfield{author}{\bibinfo{person}{Paul Denny}, \bibinfo{person}{Juho
  Leinonen}, \bibinfo{person}{James Prather}, \bibinfo{person}{Andrew
  Luxton-Reilly}, \bibinfo{person}{Thezyrie Amarouche},
  \bibinfo{person}{Brett~A Becker}, {and} \bibinfo{person}{Brent~N Reeves}.}
  \bibinfo{year}{2023}\natexlab{b}.
\newblock \showarticletitle{Promptly: Using Prompt Problems to Teach Learners
  How to Effectively Utilize AI Code Generators}.
\newblock \bibinfo{journal}{\emph{arXiv preprint arXiv:2307.16364}}
  (\bibinfo{year}{2023}).
\newblock


\bibitem[Denny et~al\mbox{.}(2012)]%
        {denny2012all}
\bibfield{author}{\bibinfo{person}{Paul Denny}, \bibinfo{person}{Andrew
  Luxton-Reilly}, {and} \bibinfo{person}{Ewan Tempero}.}
  \bibinfo{year}{2012}\natexlab{}.
\newblock \showarticletitle{All Syntax Errors Are Not Equal}. In
  \bibinfo{booktitle}{\emph{Proc. of the 17th ACM Annual Conf. on Innovation
  and Technology in Computer Science Education}}. \bibinfo{publisher}{ACM},
  \bibinfo{address}{NY, NY, USA}, \bibinfo{pages}{75--80}.
\newblock
\showISBNx{978-1-4503-1246-2}


\bibitem[Denny et~al\mbox{.}(2021)]%
        {denny2021designing}
\bibfield{author}{\bibinfo{person}{Paul Denny}, \bibinfo{person}{James
  Prather}, \bibinfo{person}{Brett~A Becker}, \bibinfo{person}{Catherine
  Mooney}, \bibinfo{person}{John Homer}, \bibinfo{person}{Zachary~C Albrecht},
  {and} \bibinfo{person}{Garrett~B Powell}.} \bibinfo{year}{2021}\natexlab{}.
\newblock \showarticletitle{On Designing Programming Error Messages for
  Novices: Readability and Its Constituent Factors}. In
  \bibinfo{booktitle}{\emph{Proc. of the 2021 CHI Conf. on Human Factors in
  Computing Systems}}. \bibinfo{pages}{1--15}.
\newblock


\bibitem[Ettles et~al\mbox{.}(2018)]%
        {ettles2018common}
\bibfield{author}{\bibinfo{person}{Andrew Ettles}, \bibinfo{person}{Andrew
  Luxton-Reilly}, {and} \bibinfo{person}{Paul Denny}.}
  \bibinfo{year}{2018}\natexlab{}.
\newblock \showarticletitle{Common Logic Errors Made by Novice Programmers}. In
  \bibinfo{booktitle}{\emph{Proc. of the 20th Australasian Computing Education
  Conf.}} \bibinfo{publisher}{ACM}, \bibinfo{address}{New York, NY, USA},
  \bibinfo{pages}{83–89}.
\newblock
\showISBNx{9781450363402}


\bibitem[Fan et~al\mbox{.}(2023)]%
        {fan2023automated}
\bibfield{author}{\bibinfo{person}{Zhiyu Fan}, \bibinfo{person}{Xiang Gao},
  \bibinfo{person}{Martin Mirchev}, \bibinfo{person}{Abhik Roychoudhury}, {and}
  \bibinfo{person}{Shin~Hwei Tan}.} \bibinfo{year}{2023}\natexlab{}.
\newblock \showarticletitle{Automated repair of programs from large language
  models}. In \bibinfo{booktitle}{\emph{2023 IEEE/ACM 45th Int. Conf. on
  Software Engineering (ICSE)}}. IEEE, \bibinfo{pages}{1469--1481}.
\newblock


\bibitem[Finnie-Ansley et~al\mbox{.}(2022)]%
        {Finnie-Ansley2022Robots}
\bibfield{author}{\bibinfo{person}{James Finnie-Ansley}, \bibinfo{person}{Paul
  Denny}, \bibinfo{person}{Brett~A. Becker}, \bibinfo{person}{Andrew
  Luxton-Reilly}, {and} \bibinfo{person}{James Prather}.}
  \bibinfo{year}{2022}\natexlab{}.
\newblock \showarticletitle{The Robots Are Coming: Exploring the Implications
  of OpenAI Codex on Introductory Programming}. In
  \bibinfo{booktitle}{\emph{Australasian Computing Education Conf.}}
  \bibinfo{publisher}{ACM}, \bibinfo{address}{New York, NY, USA},
  \bibinfo{pages}{10–19}.
\newblock
\showISBNx{9781450396431}


\bibitem[Fitzgerald et~al\mbox{.}(2008)]%
        {fitzgerald2008debugging}
\bibfield{author}{\bibinfo{person}{Sue Fitzgerald}, \bibinfo{person}{Gary
  Lewandowski}, \bibinfo{person}{Renee McCauley}, \bibinfo{person}{Laurie
  Murphy}, \bibinfo{person}{Beth Simon}, \bibinfo{person}{Lynda Thomas}, {and}
  \bibinfo{person}{Carol Zander}.} \bibinfo{year}{2008}\natexlab{}.
\newblock \showarticletitle{Debugging: finding, fixing and flailing, a
  multi-institutional study of novice debuggers}.
\newblock \bibinfo{journal}{\emph{Computer Science Education}}
  \bibinfo{volume}{18}, \bibinfo{number}{2} (\bibinfo{year}{2008}),
  \bibinfo{pages}{93--116}.
\newblock


\bibitem[Glassman et~al\mbox{.}(2015)]%
        {glassman2015overcode}
\bibfield{author}{\bibinfo{person}{Elena~L Glassman}, \bibinfo{person}{Jeremy
  Scott}, \bibinfo{person}{Rishabh Singh}, \bibinfo{person}{Philip~J Guo},
  {and} \bibinfo{person}{Robert~C Miller}.} \bibinfo{year}{2015}\natexlab{}.
\newblock \showarticletitle{{OverCode}: Visualizing variation in student
  solutions to programming problems at scale}.
\newblock \bibinfo{journal}{\emph{ACM Trans. on Computer-Human Interaction
  (TOCHI)}} \bibinfo{volume}{22}, \bibinfo{number}{2} (\bibinfo{year}{2015}),
  \bibinfo{pages}{1--35}.
\newblock


\bibitem[Heinonen et~al\mbox{.}(2023)]%
        {heinonen2023synthesizing}
\bibfield{author}{\bibinfo{person}{Ava Heinonen}, \bibinfo{person}{Bettina
  Lehtel{\"a}}, \bibinfo{person}{Arto Hellas}, {and} \bibinfo{person}{Fabian
  Fagerholm}.} \bibinfo{year}{2023}\natexlab{}.
\newblock \showarticletitle{Synthesizing research on programmers’ mental
  models of programs, tasks and concepts—A systematic literature review}.
\newblock \bibinfo{journal}{\emph{Information and Software Technology}}
  (\bibinfo{year}{2023}), \bibinfo{pages}{107300}.
\newblock


\bibitem[Hellas et~al\mbox{.}(2023)]%
        {hellas2023exploring}
\bibfield{author}{\bibinfo{person}{Arto Hellas}, \bibinfo{person}{Juho
  Leinonen}, \bibinfo{person}{Sami Sarsa}, \bibinfo{person}{Charles Koutcheme},
  \bibinfo{person}{Lilja Kujanp{\"a}{\"a}}, {and} \bibinfo{person}{Juha
  Sorva}.} \bibinfo{year}{2023}\natexlab{}.
\newblock \showarticletitle{Exploring the Responses of Large Language Models to
  Beginner Programmers' Help Requests}.
\newblock \bibinfo{journal}{\emph{arXiv preprint arXiv:2306.05715}}
  (\bibinfo{year}{2023}).
\newblock


\bibitem[Hou et~al\mbox{.}(2023)]%
        {hou2023more}
\bibfield{author}{\bibinfo{person}{Irene Hou}, \bibinfo{person}{Owen Man},
  \bibinfo{person}{Sophie Mettille}, \bibinfo{person}{Sebastian Gutierrez},
  \bibinfo{person}{Kenneth Angelikas}, {and} \bibinfo{person}{Stephen
  MacNeil}.} \bibinfo{year}{2023}\natexlab{}.
\newblock \showarticletitle{More Robots are Coming: Large Multimodal Models
  (ChatGPT) can Solve Visually Diverse Images of Parsons Problems}.
\newblock \bibinfo{journal}{\emph{arXiv preprint arXiv:2311.04926}}
  (\bibinfo{year}{2023}).
\newblock
\urldef\tempurl%
\url{https://doi.org/10.48550/arXiv.2311.04926}
\showDOI{\tempurl}


\bibitem[Hou et~al\mbox{.}(2024)]%
        {hou2023effects}
\bibfield{author}{\bibinfo{person}{Irene Hou}, \bibinfo{person}{Sophia
  Mettille}, \bibinfo{person}{Owen Man}, \bibinfo{person}{Zhuo Li},
  \bibinfo{person}{Cynthia Zastudil}, {and} \bibinfo{person}{Stephen MacNeil}.}
  \bibinfo{year}{2024}\natexlab{}.
\newblock \showarticletitle{The Effects of Generative AI on Introductory
  Students’ Help-Seeking Preferences}. In
  \bibinfo{booktitle}{\emph{Australasian Computing Education Conference}}
  \emph{(\bibinfo{series}{ACM ACE '24})}.
\newblock


\bibitem[Jiang et~al\mbox{.}(2023)]%
        {jiang2023impact}
\bibfield{author}{\bibinfo{person}{Nan Jiang}, \bibinfo{person}{Kevin Liu},
  \bibinfo{person}{Thibaud Lutellier}, {and} \bibinfo{person}{Lin Tan}.}
  \bibinfo{year}{2023}\natexlab{}.
\newblock \showarticletitle{Impact of code language models on automated program
  repair}.
\newblock \bibinfo{journal}{\emph{arXiv preprint arXiv:2302.05020}}
  (\bibinfo{year}{2023}).
\newblock


\bibitem[Johnson et~al\mbox{.}(1983)]%
        {johnson1983bug}
\bibfield{author}{\bibinfo{person}{W.~Lewis Johnson}, \bibinfo{person}{Elliot
  Soloway}, \bibinfo{person}{Benjamin Cutler}, {and} \bibinfo{person}{Steven
  Draper}.} \bibinfo{year}{1983}\natexlab{}.
\newblock \bibinfo{booktitle}{\emph{{Bug Catalogue: I}}}.
\newblock \bibinfo{type}{{T}echnical {R}eport}. \bibinfo{institution}{Yale
  University, YaleU/CSD/RR \#286}.
\newblock


\bibitem[Kabir et~al\mbox{.}(2023)]%
        {kabir2023answers}
\bibfield{author}{\bibinfo{person}{Samia Kabir}, \bibinfo{person}{David~N
  Udo-Imeh}, \bibinfo{person}{Bonan Kou}, {and} \bibinfo{person}{Tianyi
  Zhang}.} \bibinfo{year}{2023}\natexlab{}.
\newblock \showarticletitle{Who Answers It Better? An In-Depth Analysis of
  ChatGPT and Stack Overflow Answers to Software Engineering Questions}.
\newblock \bibinfo{journal}{\emph{arXiv preprint arXiv:2308.02312}}
  (\bibinfo{year}{2023}).
\newblock


\bibitem[Kohn(2019)]%
        {kohn2019error}
\bibfield{author}{\bibinfo{person}{Tobias Kohn}.}
  \bibinfo{year}{2019}\natexlab{}.
\newblock \showarticletitle{The error behind the message: Finding the cause of
  error messages in python}. In \bibinfo{booktitle}{\emph{Proc. of the 50th ACM
  Technical Symp. on Computer Science Education}}. \bibinfo{pages}{524--530}.
\newblock


\bibitem[Koivisto and Hellas(2022)]%
        {koivisto2022evaluating}
\bibfield{author}{\bibinfo{person}{Teemu Koivisto} {and} \bibinfo{person}{Arto
  Hellas}.} \bibinfo{year}{2022}\natexlab{}.
\newblock \showarticletitle{Evaluating CodeClusters for Effectively Providing
  Feedback on Code Submissions}. In \bibinfo{booktitle}{\emph{2022 IEEE
  Frontiers in Education Conf. (FIE)}}. IEEE, \bibinfo{pages}{1--9}.
\newblock


\bibitem[Koutcheme et~al\mbox{.}(2023)]%
        {koutcheme2023automated}
\bibfield{author}{\bibinfo{person}{Charles Koutcheme}, \bibinfo{person}{Sami
  Sarsa}, \bibinfo{person}{Juho Leinonen}, \bibinfo{person}{Arto Hellas}, {and}
  \bibinfo{person}{Paul Denny}.} \bibinfo{year}{2023}\natexlab{}.
\newblock \showarticletitle{Automated Program Repair Using Generative Models
  for Code Infilling}. In \bibinfo{booktitle}{\emph{Int. Conf. on Artificial
  Intelligence in Education}}. Springer, \bibinfo{pages}{798--803}.
\newblock


\bibitem[Lau and Guo(2023)]%
        {lau2023from}
\bibfield{author}{\bibinfo{person}{Sam Lau} {and} \bibinfo{person}{Philip~J.
  Guo}.} \bibinfo{year}{2023}\natexlab{}.
\newblock \showarticletitle{From `Ban It Till We Understand It' to "Resistance
  is Futile": How University Programming Instructors Plan to Adapt as More
  Students Use AI Code Generation and Explanation Tools such as ChatGPT and
  GitHub Copilot}. In \bibinfo{booktitle}{\emph{In Proc. of the 2023 ACM Conf.
  on Int. Computing Education Research V.1 (ICER ’23 V1)}}.
  \bibinfo{publisher}{ACM}.
\newblock
\urldef\tempurl%
\url{https://doi.org/10.1145/3568813.3600138}
\showDOI{\tempurl}


\bibitem[Leinonen et~al\mbox{.}(2023a)]%
        {leinonen2023comparing}
\bibfield{author}{\bibinfo{person}{Juho Leinonen}, \bibinfo{person}{Paul
  Denny}, \bibinfo{person}{Stephen MacNeil}, \bibinfo{person}{Sami Sarsa},
  \bibinfo{person}{Seth Bernstein}, \bibinfo{person}{Joanne Kim},
  \bibinfo{person}{Andrew Tran}, {and} \bibinfo{person}{Arto Hellas}.}
  \bibinfo{year}{2023}\natexlab{a}.
\newblock \showarticletitle{Comparing Code Explanations Created by Students and
  Large Language Models}.
\newblock \bibinfo{journal}{\emph{arXiv preprint arXiv:2304.03938}}
  (\bibinfo{year}{2023}).
\newblock


\bibitem[Leinonen et~al\mbox{.}(2023b)]%
        {leinonen2023using}
\bibfield{author}{\bibinfo{person}{Juho Leinonen}, \bibinfo{person}{Arto
  Hellas}, \bibinfo{person}{Sami Sarsa}, \bibinfo{person}{Brent Reeves},
  \bibinfo{person}{Paul Denny}, \bibinfo{person}{James Prather}, {and}
  \bibinfo{person}{Brett~A Becker}.} \bibinfo{year}{2023}\natexlab{b}.
\newblock \showarticletitle{Using large language models to enhance programming
  error messages}. In \bibinfo{booktitle}{\emph{Proc. of the 54th ACM Technical
  Symp. on Computer Science Education V. 1}}. \bibinfo{pages}{563--569}.
\newblock


\bibitem[Lister et~al\mbox{.}(2004)]%
        {lister2004multi}
\bibfield{author}{\bibinfo{person}{Raymond Lister},
  \bibinfo{person}{Elizabeth~S Adams}, \bibinfo{person}{Sue Fitzgerald},
  \bibinfo{person}{William Fone}, \bibinfo{person}{John Hamer},
  \bibinfo{person}{Morten Lindholm}, \bibinfo{person}{Robert McCartney},
  \bibinfo{person}{Jan~Erik Mostr{\"o}m}, \bibinfo{person}{Kate Sanders},
  \bibinfo{person}{Otto Sepp{\"a}l{\"a}}, {et~al\mbox{.}}}
  \bibinfo{year}{2004}\natexlab{}.
\newblock \showarticletitle{A multi-national study of reading and tracing
  skills in novice programmers}.
\newblock \bibinfo{journal}{\emph{ACM SIGCSE Bulletin}} \bibinfo{volume}{36},
  \bibinfo{number}{4} (\bibinfo{year}{2004}), \bibinfo{pages}{119--150}.
\newblock


\bibitem[Lister et~al\mbox{.}(2009)]%
        {lister2009further}
\bibfield{author}{\bibinfo{person}{Raymond Lister}, \bibinfo{person}{Colin
  Fidge}, {and} \bibinfo{person}{Donna Teague}.}
  \bibinfo{year}{2009}\natexlab{}.
\newblock \showarticletitle{Further Evidence of a Relationship between
  Explaining, Tracing and Writing Skills in Introductory Programming}.
\newblock \bibinfo{journal}{\emph{SIGCSE Bull.}} \bibinfo{volume}{41},
  \bibinfo{number}{3} (\bibinfo{year}{2009}), \bibinfo{pages}{161–165}.
\newblock
\showISSN{0097-8418}


\bibitem[MacNeil et~al\mbox{.}(2023a)]%
        {macneil2023implications}
\bibfield{author}{\bibinfo{person}{Stephen MacNeil}, \bibinfo{person}{Joanne
  Kim}, \bibinfo{person}{Juho Leinonen}, \bibinfo{person}{Paul Denny},
  \bibinfo{person}{Seth Bernstein}, \bibinfo{person}{Brett~A. Becker},
  \bibinfo{person}{Michel Wermelinger}, \bibinfo{person}{Arto Hellas},
  \bibinfo{person}{Andrew Tran}, \bibinfo{person}{Sami Sarsa},
  \bibinfo{person}{James Prather}, {and} \bibinfo{person}{Viraj Kumar}.}
  \bibinfo{year}{2023}\natexlab{a}.
\newblock \showarticletitle{The Implications of Large Language Models for CS
  Teachers and Students}. In \bibinfo{booktitle}{\emph{Proc. of the 54th ACM
  Technical Symp. on Computer Science Education V. 2}}.
  \bibinfo{publisher}{ACM}, \bibinfo{address}{New York, NY, USA},
  \bibinfo{pages}{1255}.
\newblock
\showISBNx{9781450394338}


\bibitem[MacNeil et~al\mbox{.}(2023b)]%
        {macneil2023experiences}
\bibfield{author}{\bibinfo{person}{Stephen MacNeil}, \bibinfo{person}{Andrew
  Tran}, \bibinfo{person}{Arto Hellas}, \bibinfo{person}{Joanne Kim},
  \bibinfo{person}{Sami Sarsa}, \bibinfo{person}{Paul Denny},
  \bibinfo{person}{Seth Bernstein}, {and} \bibinfo{person}{Juho Leinonen}.}
  \bibinfo{year}{2023}\natexlab{b}.
\newblock \showarticletitle{Experiences from Using Code Explanations Generated
  by Large Language Models in a Web Software Development E-Book}. In
  \bibinfo{booktitle}{\emph{Proc. SIGCSE'23}}. \bibinfo{publisher}{ACM},
  \bibinfo{numpages}{6}~pages.
\newblock


\bibitem[MacNeil et~al\mbox{.}(2023c)]%
        {macneil2022automatically}
\bibfield{author}{\bibinfo{person}{Stephen MacNeil}, \bibinfo{person}{Andrew
  Tran}, \bibinfo{person}{Juho Leinonen}, \bibinfo{person}{Paul Denny},
  \bibinfo{person}{Joanne Kim}, \bibinfo{person}{Arto Hellas},
  \bibinfo{person}{Seth Bernstein}, {and} \bibinfo{person}{Sami Sarsa}.}
  \bibinfo{year}{2023}\natexlab{c}.
\newblock \showarticletitle{Automatically Generating CS Learning Materials with
  Large Language Models}. In \bibinfo{booktitle}{\emph{Proc. of the 54th ACM
  Technical Symp. on Computer Science Education V. 2}}.
  \bibinfo{publisher}{ACM}, \bibinfo{address}{New York, NY, USA},
  \bibinfo{pages}{1176}.
\newblock
\showISBNx{9781450394338}


\bibitem[Malysheva and Kelleher(2020)]%
        {malysheva2020bugs}
\bibfield{author}{\bibinfo{person}{Yana Malysheva} {and}
  \bibinfo{person}{Caitlin Kelleher}.} \bibinfo{year}{2020}\natexlab{}.
\newblock \showarticletitle{Bugs as Features: Describing Patterns in Student
  Code through a Classification of Bugs}. In \bibinfo{booktitle}{\emph{Extended
  Abstracts of the 2020 CHI Conf. on Human Factors in Computing Systems}}.
  \bibinfo{publisher}{ACM}, \bibinfo{address}{New York, NY, USA},
  \bibinfo{pages}{1–7}.
\newblock
\showISBNx{9781450368193}


\bibitem[McCall and K{\"o}lling(2019)]%
        {mccall2019new}
\bibfield{author}{\bibinfo{person}{Davin McCall} {and} \bibinfo{person}{Michael
  K{\"o}lling}.} \bibinfo{year}{2019}\natexlab{}.
\newblock \showarticletitle{A new look at novice programmer errors}.
\newblock \bibinfo{journal}{\emph{ACM Trans. on Computing Education (TOCE)}}
  \bibinfo{volume}{19}, \bibinfo{number}{4} (\bibinfo{year}{2019}),
  \bibinfo{pages}{1--30}.
\newblock


\bibitem[McCracken et~al\mbox{.}(2001)]%
        {mccracken2001multi}
\bibfield{author}{\bibinfo{person}{Michael McCracken}, \bibinfo{person}{Vicki
  Almstrum}, \bibinfo{person}{Danny Diaz}, \bibinfo{person}{Mark Guzdial},
  \bibinfo{person}{Dianne Hagan}, \bibinfo{person}{Yifat Ben-David Kolikant},
  \bibinfo{person}{Cary Laxer}, \bibinfo{person}{Lynda Thomas},
  \bibinfo{person}{Ian Utting}, {and} \bibinfo{person}{Tadeusz Wilusz}.}
  \bibinfo{year}{2001}\natexlab{}.
\newblock \showarticletitle{A multi-national, multi-institutional study of
  assessment of programming skills of first-year CS students}.
\newblock In \bibinfo{booktitle}{\emph{Working group reports from ITiCSE on
  Innovation and technology in computer science education}}.
  \bibinfo{pages}{125--180}.
\newblock


\bibitem[Murphy et~al\mbox{.}(2008)]%
        {murphy2008debugging}
\bibfield{author}{\bibinfo{person}{Laurie Murphy}, \bibinfo{person}{Gary
  Lewandowski}, \bibinfo{person}{Ren{\'e}e McCauley}, \bibinfo{person}{Beth
  Simon}, \bibinfo{person}{Lynda Thomas}, {and} \bibinfo{person}{Carol
  Zander}.} \bibinfo{year}{2008}\natexlab{}.
\newblock \showarticletitle{Debugging: the good, the bad, and the quirky--a
  qualitative analysis of novices' strategies}. In
  \bibinfo{booktitle}{\emph{ACM SIGCSE Bulletin}}, Vol.~\bibinfo{volume}{40}.
  ACM.
\newblock


\bibitem[Prather et~al\mbox{.}(2023a)]%
        {prather2023transformed}
\bibfield{author}{\bibinfo{person}{James Prather}, \bibinfo{person}{Paul
  Denny}, \bibinfo{person}{Juho Leinonen}, \bibinfo{person}{Brett~A Becker},
  \bibinfo{person}{Ibrahim Albluwi}, \bibinfo{person}{Michael~E Caspersen},
  \bibinfo{person}{Michelle Craig}, \bibinfo{person}{Hieke Keuning},
  \bibinfo{person}{Natalie Kiesler}, \bibinfo{person}{Tobias Kohn},
  {et~al\mbox{.}}} \bibinfo{year}{2023}\natexlab{a}.
\newblock \showarticletitle{Transformed by Transformers: Navigating the AI
  Coding Revolution for Computing Education: An ITiCSE Working Group Conducted
  by Humans}. In \bibinfo{booktitle}{\emph{Proc. of the 2023 Conf. on
  Innovation and Technology in Computer Science Education V. 2}}.
  \bibinfo{pages}{561--562}.
\newblock


\bibitem[Prather et~al\mbox{.}(2023b)]%
        {prather2023robots}
\bibfield{author}{\bibinfo{person}{James Prather}, \bibinfo{person}{Paul
  Denny}, \bibinfo{person}{Juho Leinonen}, \bibinfo{person}{Brett~A Becker},
  \bibinfo{person}{Ibrahim Albluwi}, \bibinfo{person}{Michelle Craig},
  \bibinfo{person}{Hieke Keuning}, \bibinfo{person}{Natalie Kiesler},
  \bibinfo{person}{Tobias Kohn}, \bibinfo{person}{Andrew Luxton-Reilly},
  {et~al\mbox{.}}} \bibinfo{year}{2023}\natexlab{b}.
\newblock \showarticletitle{The robots are here: Navigating the generative ai
  revolution in computing education}.
\newblock \bibinfo{journal}{\emph{arXiv preprint arXiv:2310.00658}}
  (\bibinfo{year}{2023}).
\newblock


\bibitem[Puryear and Sprint(2022)]%
        {puryear2022github}
\bibfield{author}{\bibinfo{person}{Ben Puryear} {and} \bibinfo{person}{Gina
  Sprint}.} \bibinfo{year}{2022}\natexlab{}.
\newblock \showarticletitle{Github copilot in the classroom: learning to code
  with AI assistance}.
\newblock \bibinfo{journal}{\emph{J. of Computing Sciences in Colleges}}
  \bibinfo{volume}{38}, \bibinfo{number}{1} (\bibinfo{year}{2022}),
  \bibinfo{pages}{37--47}.
\newblock


\bibitem[Sarsa et~al\mbox{.}(2022)]%
        {sarsa2022automatic}
\bibfield{author}{\bibinfo{person}{Sami Sarsa}, \bibinfo{person}{Paul Denny},
  \bibinfo{person}{Arto Hellas}, {and} \bibinfo{person}{Juho Leinonen}.}
  \bibinfo{year}{2022}\natexlab{}.
\newblock \showarticletitle{Automatic Generation of Programming Exercises and
  Code Explanations Using Large Language Models}. In
  \bibinfo{booktitle}{\emph{Proc. of the 2022 ACM Conf. on Int. Computing
  Education Research - Volume 1}}. \bibinfo{publisher}{ACM},
  \bibinfo{pages}{27–43}.
\newblock
\showISBNx{9781450391948}


\bibitem[Savelka et~al\mbox{.}(2023a)]%
        {savelka2023thrilled}
\bibfield{author}{\bibinfo{person}{Jaromir Savelka}, \bibinfo{person}{Arav
  Agarwal}, \bibinfo{person}{Marshall An}, \bibinfo{person}{Chris Bogart},
  {and} \bibinfo{person}{Majd Sakr}.} \bibinfo{year}{2023}\natexlab{a}.
\newblock \showarticletitle{Thrilled by Your Progress! Large Language Models
  (GPT-4) No Longer Struggle to Pass Assessments in Higher Education
  Programming Courses}.
\newblock \bibinfo{journal}{\emph{The 19th ACM Conference on International
  Computing Education Research (ICER)}} (\bibinfo{year}{2023}).
\newblock


\bibitem[Savelka et~al\mbox{.}(2023b)]%
        {savelka2023large}
\bibfield{author}{\bibinfo{person}{Jaromir Savelka}, \bibinfo{person}{Arav
  Agarwal}, \bibinfo{person}{Christopher Bogart}, {and} \bibinfo{person}{Majd
  Sakr}.} \bibinfo{year}{2023}\natexlab{b}.
\newblock \bibinfo{title}{Large Language Models (GPT) Struggle to Answer
  Multiple-Choice Questions about Code}.
\newblock
\newblock
\showeprint[arxiv]{2303.08033}~[cs.CL]


\bibitem[Savelka et~al\mbox{.}(2023c)]%
        {savelka2023can}
\bibfield{author}{\bibinfo{person}{Jaromir Savelka}, \bibinfo{person}{Arav
  Agarwal}, \bibinfo{person}{Christopher Bogart}, \bibinfo{person}{Yifan Song},
  {and} \bibinfo{person}{Majd Sakr}.} \bibinfo{year}{2023}\natexlab{c}.
\newblock \showarticletitle{Can Generative Pre-trained Transformers (GPT) Pass
  Assessments in Higher Education Programming Courses?}
\newblock \bibinfo{journal}{\emph{arXiv preprint arXiv:2303.09325}}
  (\bibinfo{year}{2023}).
\newblock


\bibitem[Sepp{\"a}l{\"a} et~al\mbox{.}(2015)]%
        {seppala2015we}
\bibfield{author}{\bibinfo{person}{Otto Sepp{\"a}l{\"a}},
  \bibinfo{person}{Petri Ihantola}, \bibinfo{person}{Essi Isohanni},
  \bibinfo{person}{Juha Sorva}, {and} \bibinfo{person}{Arto Vihavainen}.}
  \bibinfo{year}{2015}\natexlab{}.
\newblock \showarticletitle{Do we know how difficult the rainfall problem is?}.
  In \bibinfo{booktitle}{\emph{Proc. of the 15th Koli Calling Conf. on
  Computing Education Research}}. \bibinfo{pages}{87--96}.
\newblock


\bibitem[Si et~al\mbox{.}(2023)]%
        {si2023prompting}
\bibfield{author}{\bibinfo{person}{Chenglei Si}, \bibinfo{person}{Zhe Gan},
  \bibinfo{person}{Zhengyuan Yang}, \bibinfo{person}{Shuohang Wang},
  \bibinfo{person}{Jianfeng Wang}, \bibinfo{person}{Jordan Boyd-Graber}, {and}
  \bibinfo{person}{Lijuan Wang}.} \bibinfo{year}{2023}\natexlab{}.
\newblock \bibinfo{title}{Prompting GPT-3 To Be Reliable}.
\newblock
\newblock
\showeprint[arxiv]{2210.09150}~[cs.CL]


\bibitem[Smith and Rixner(2019)]%
        {smith2019error}
\bibfield{author}{\bibinfo{person}{Rebecca Smith} {and} \bibinfo{person}{Scott
  Rixner}.} \bibinfo{year}{2019}\natexlab{}.
\newblock \showarticletitle{The error landscape: Characterizing the mistakes of
  novice programmers}. In \bibinfo{booktitle}{\emph{Proc. of the 50th ACM
  technical Symp. on computer science education}}. \bibinfo{pages}{538--544}.
\newblock


\bibitem[Soloway et~al\mbox{.}(1983)]%
        {soloway1983cognitive}
\bibfield{author}{\bibinfo{person}{Elliot Soloway}, \bibinfo{person}{Jeffrey~G.
  Bonar}, {and} \bibinfo{person}{Kate Ehrlich}.}
  \bibinfo{year}{1983}\natexlab{}.
\newblock \showarticletitle{{Cognitive strategies and looping constructs: An
  empirical study}}.
\newblock \bibinfo{journal}{\emph{Commun. ACM}} \bibinfo{volume}{26},
  \bibinfo{number}{11} (\bibinfo{year}{1983}), \bibinfo{pages}{853--860}.
\newblock


\bibitem[Soloway et~al\mbox{.}(1982)]%
        {soloway1982what}
\bibfield{author}{\bibinfo{person}{Elliot Soloway}, \bibinfo{person}{Kate
  Ehrlich}, \bibinfo{person}{Jeffrey~G. Bonar}, {and} \bibinfo{person}{Judith
  Greenspan}.} \bibinfo{year}{1982}\natexlab{}.
\newblock \showarticletitle{{What do novices know about programming?}}
\newblock In \bibinfo{booktitle}{\emph{Directions in Human--Computer
  Interactions}}. Vol.~\bibinfo{volume}{6}. \bibinfo{publisher}{Ablex
  Publishing}, \bibinfo{pages}{27--54}.
\newblock


\bibitem[Spohrer and Soloway(1986)]%
        {spohrer1986novice}
\bibfield{author}{\bibinfo{person}{James~C Spohrer} {and}
  \bibinfo{person}{Elliot Soloway}.} \bibinfo{year}{1986}\natexlab{}.
\newblock \showarticletitle{Novice mistakes: Are the folk wisdoms correct?}
\newblock \bibinfo{journal}{\emph{Commun. ACM}} \bibinfo{volume}{29},
  \bibinfo{number}{7} (\bibinfo{year}{1986}), \bibinfo{pages}{624--632}.
\newblock


\bibitem[Teebagy et~al\mbox{.}(2023)]%
        {teebagy2023improved}
\bibfield{author}{\bibinfo{person}{Sean Teebagy}, \bibinfo{person}{Lauren
  Colwell}, \bibinfo{person}{Emma Wood}, \bibinfo{person}{Antonio Yaghy}, {and}
  \bibinfo{person}{Misha Faustina}.} \bibinfo{year}{2023}\natexlab{}.
\newblock \showarticletitle{Improved performance of ChatGPT-4 on the OKAP exam:
  A comparative study with ChatGPT-3.5}.
\newblock \bibinfo{journal}{\emph{medRxiv}} (\bibinfo{year}{2023}),
  \bibinfo{pages}{2023--04}.
\newblock


\bibitem[Tran et~al\mbox{.}(2023)]%
        {tran2023using}
\bibfield{author}{\bibinfo{person}{Andrew Tran}, \bibinfo{person}{Linxuan Li},
  \bibinfo{person}{Egi Rama}, \bibinfo{person}{Kenneth Angelikas}, {and}
  \bibinfo{person}{Stephen MacNeil}.} \bibinfo{year}{2023}\natexlab{}.
\newblock \showarticletitle{Using Large Language Models to Automatically
  Identify Programming Concepts in Code Snippets}. In
  \bibinfo{booktitle}{\emph{Proc. of the 2023 ACM Conf. on Int. Computing
  Education Research - Volume 2}}, Vol.~\bibinfo{volume}{1}.
  \bibinfo{publisher}{ACM}, \bibinfo{pages}{563--569}.
\newblock


\bibitem[Vainio and Sajaniemi(2007)]%
        {vainio2007factors}
\bibfield{author}{\bibinfo{person}{Vesa Vainio} {and} \bibinfo{person}{Jorma
  Sajaniemi}.} \bibinfo{year}{2007}\natexlab{}.
\newblock \showarticletitle{Factors in novice programmers' poor tracing
  skills}.
\newblock \bibinfo{journal}{\emph{ACM SIGCSE Bulletin}} \bibinfo{volume}{39},
  \bibinfo{number}{3} (\bibinfo{year}{2007}), \bibinfo{pages}{236--240}.
\newblock


\bibitem[Wermelinger(2023)]%
        {wermelinger2023using}
\bibfield{author}{\bibinfo{person}{Michel Wermelinger}.}
  \bibinfo{year}{2023}\natexlab{}.
\newblock \showarticletitle{Using GitHub Copilot to Solve Simple Programming
  Problems}. In \bibinfo{booktitle}{\emph{Proc. of the 54th ACM Technical Symp.
  on Computer Science Education V. 1}}. \bibinfo{publisher}{ACM}.
\newblock


\bibitem[Whalley et~al\mbox{.}(2007)]%
        {whalley2007many}
\bibfield{author}{\bibinfo{person}{Jacqueline Whalley}, \bibinfo{person}{Tony
  Clear}, {and} \bibinfo{person}{RF Lister}.} \bibinfo{year}{2007}\natexlab{}.
\newblock \showarticletitle{The many ways of the BRACElet project}.
\newblock \bibinfo{journal}{\emph{Bull. of Applied Computing and Information
  Technology}} (\bibinfo{year}{2007}).
\newblock


\bibitem[Whalley et~al\mbox{.}(2021)]%
        {whalley2021novice}
\bibfield{author}{\bibinfo{person}{Jacqueline Whalley}, \bibinfo{person}{Amber
  Settle}, {and} \bibinfo{person}{Andrew Luxton-Reilly}.}
  \bibinfo{year}{2021}\natexlab{}.
\newblock \showarticletitle{Novice reflections on debugging}. In
  \bibinfo{booktitle}{\emph{Proc. of the 52nd ACM Technical Symp. on Computer
  Science Education}}. \bibinfo{pages}{73--79}.
\newblock


\bibitem[Whalley et~al\mbox{.}(2006)]%
        {whalley2006australasian}
\bibfield{author}{\bibinfo{person}{Jacqueline~L. Whalley},
  \bibinfo{person}{Raymond Lister}, \bibinfo{person}{Errol Thompson},
  \bibinfo{person}{Tony Clear}, \bibinfo{person}{Phil Robbins},
  \bibinfo{person}{P.~K.~Ajith Kumar}, {and} \bibinfo{person}{Christine
  Prasad}.} \bibinfo{year}{2006}\natexlab{}.
\newblock \showarticletitle{An Australasian Study of Reading and Comprehension
  Skills in Novice Programmers, Using the Bloom and SOLO Taxonomies}. In
  \bibinfo{booktitle}{\emph{Proc. of the 8th Australasian Conf. on Computing
  Education - Volume 52}}. \bibinfo{publisher}{Australian Computer Society,
  Inc.}, \bibinfo{address}{AUS}, \bibinfo{pages}{243–252}.
\newblock
\showISBNx{1920682341}


\bibitem[Winikoff(2014)]%
        {winikoff2014novice}
\bibfield{author}{\bibinfo{person}{Michael Winikoff}.}
  \bibinfo{year}{2014}\natexlab{}.
\newblock \showarticletitle{Novice programmers' faults \& failures in GOAL
  programs}. In \bibinfo{booktitle}{\emph{Proc. of the 2014 Int. Conf. on
  Autonomous agents and multi-agent systems}}. \bibinfo{pages}{301--308}.
\newblock


\bibitem[Zastudil et~al\mbox{.}(2023)]%
        {zastudil2023generative}
\bibfield{author}{\bibinfo{person}{Cynthia Zastudil},
  \bibinfo{person}{Magdalena Rogalska}, \bibinfo{person}{Christine Kapp},
  \bibinfo{person}{Jennifer Vaughn}, {and} \bibinfo{person}{Stephen MacNeil}.}
  \bibinfo{year}{2023}\natexlab{}.
\newblock \showarticletitle{Generative AI in Computing Education: Perspectives
  of Students and Instructors}.
\newblock \bibinfo{journal}{\emph{arXiv preprint arXiv:2308.04309}}
  (\bibinfo{year}{2023}).
\newblock
\urldef\tempurl%
\url{https://doi.org/10.48550/arXiv.2308.04309}
\showDOI{\tempurl}


\bibitem[Zhao et~al\mbox{.}(2021)]%
        {Zhao2021CalibrateBU}
\bibfield{author}{\bibinfo{person}{Tony Zhao}, \bibinfo{person}{Eric Wallace},
  \bibinfo{person}{Shi Feng}, \bibinfo{person}{Dan Klein}, {and}
  \bibinfo{person}{Sameer Singh}.} \bibinfo{year}{2021}\natexlab{}.
\newblock \showarticletitle{Calibrate Before Use: Improving Few-Shot
  Performance of Language Models}. In \bibinfo{booktitle}{\emph{Int. Conf. on
  Machine Learning}}.
\newblock


\end{thebibliography}


\end{document}